\documentclass[10pt,letterpaper,twocolumn]{article} 

\usepackage[hyphens]{url}
\usepackage[usenames,dvipsnames]{xcolor}

\usepackage[]{hyperref}
\hypersetup{
    colorlinks=true,
    breaklinks=true
}
\usepackage[hyphens]{url}

\usepackage[hyphenbreaks]{breakurl}

\usepackage[10pt,nocopyright]{sigmin}
\usepackage[square,comma,numbers,sort&compress]{natbib}
\usepackage{authblk}

\usepackage{silence}
\usepackage{amsmath,amsopn,amssymb}
\usepackage{subfigure}
\usepackage{endnotes,microtype,xspace,graphicx,fancyvrb,multirow}
\usepackage{booktabs}
\usepackage{array,underscore,relsize}
\usepackage[T1]{fontenc}
\usepackage{times}
\usepackage{fancyhdr,lastpage}
\usepackage{enumitem}
\usepackage[labelfont=bf,font=small,skip=5pt]{caption}
\usepackage{fp}
\usepackage{siunitx}

\usepackage{balance}

\usepackage{geometry}

\usepackage{cleveref}

\newcommand{\sys}{\textsc{Pixiu}\xspace}
\newcommand{\tstep}{trust-$\lambda$\xspace}

\usepackage{mfirstuc}
\usepackage{ulem}

\fvset{fontsize=\scriptsize,xleftmargin=8pt,numbers=left,numbersep=5pt}

\def\Snospace~{\S{}}

\newcommand{\x}{$\times$\xspace}

\newif\ifdraft\drafttrue
\newif\ifnotes\notestrue
\ifdraft\else\notesfalse\fi

\input{glyphtounicode}
\newcolumntype{R}[1]{>{\raggedleft\let\newline\\\arraybackslash\hspace{0pt}}p{#1}}

\newcommand{\includepdf}[1]{
  \includegraphics[width=\columnwidth]{#1}
}

\newcommand{\squishlist}{
\begin{itemize}[noitemsep,nolistsep]
  \setlength{\itemsep}{-0pt}
}
\newcommand{\squishend}{
  \end{itemize}
}

\usepackage{tikz}

\usepackage{xstring}
\newcommand{\PP}[1]{
\vspace{2px}
\noindent{\bf \IfEndWith{#1}{.}{#1}{#1.}}
}

\newcommand{\heading}[1]{
\vspace{1ex}
\noindent
\textbf{#1}
}

\newenvironment{myitemize}%
  {\begin{list}{\labelitemi}{\itemsep1pt \topsep2pt \parsep0.00in
  \partopsep=0pt \leftmargin1.2em}}%
  {\end{list}}

\newcommand{\CF}[1]{\xmakefirstuc{#1}}

\newcommand{\userpod}{POD\xspace}
\newcommand{\userpods}{PODs\xspace}
\newcommand{\dataconsumer}{data consumer\xspace}
\newcommand{\dataconsumers}{data consumers\xspace}
\newcommand{\executor}{executor\xspace}
\newcommand{\dexe}{decentralized executor\xspace}
\newcommand{\proof}{execution proof\xspace}

\gdef\therev{}
\gdef\thedate{}

\usepackage{usenix2019_v3_hotos}

\begin{document}

\title{Taming Distrust in the Decentralized Internet with \sys}

\ifdefined\DRAFT
 \pagestyle{fancyplain}
 \lhead{Rev.~\therev}
 \rhead{\thedate}
 \cfoot{\thepage\ of \pageref{LastPage}}
\fi

\author[1]{Yubin Xia}
\author[1]{Qingyuan Liu}
\author[2]{Cheng Tan}
\author[2]{Jing Leng}
\author[1]{Shangning Xu}
\author[1]{Binyu Zang}
\author[1]{Haibo Chen}
\affil[1]{\textit{Institute of Parallel and Distributed Systems, Shanghai Jiao Tong University}}
\affil[2]{\textit{New York University}}

\date{}
\maketitle

\begin{abstract}

Decentralized Internet is booming.
People are fascinated by its promise that users can truly own their data.
%
However,
in a decentralized Internet,
completing a task usually involves multiple nodes with mutual distrust.
Such distrust might eventually become a major obstacle for the growth of the
decentralized Internet.
In this paper, we analyze the distrust
using a simple model and highlight the properties required
to faithfully accomplish one task in a decentralized Internet.
We also introduce our draft solution---\sys,
a framework to mitigate the distrust among different nodes.
In \sys,
we design and utilize \textit{\tstep} and \textit{\dexe}
to achieve the above needed properties.

\end{abstract}

\section{Introduction}
\label{s:intro}

Surveys and reports suggest that current Internet is highly centralized~\cite{centralize1, centralize2}.
In a centralized Internet,
as users,
our (private
) data are collected
and stored in some centralized data silos, which are out of our control.
This misplacement of ownership might lead to catastrophic consequences---%
data abuse~\cite{fb},
data manipulation~\cite{yelpgoogle, weibo},
data breach~\cite{databreach} and so on.

Users should be in control of and be empowered to monetize their own data.
In order to achieve this goal, decentralized Internet has been proposed.
Though immature,
decentralized Internet is able to offer most of the functionalities
that nowadays centralized Internet provides, including
websites~\cite{zeronet}, 
online marketplace~\cite{openbazaar}, 
social network~\cite{diaspora}, 
online collaboration~\cite{graphite}, 
video sharing~\cite{dtube} 
and more~\cite{mastodon,steemit,solid}.

Different from current centralized architecture,
decentralized Internet is, by design, a peer-to-peer network
which allows individual users to truly own their data.
In a decentralized Internet,
user data are stored in \textit{\userpods}~\cite{solid} (Personal Online Datastores)
which are controlled by users themselves.
And, users can choose to run applications in their own \userpods for fun and profit.
%

However, there are trust issues in such decentralized Internet.
Instead of trusting a few giant tech companies,
a user now has to trust many (if not all) nodes in the peer-to-peer network,
which she may or may not know.
%
Indeed, a user can choose which nodes to collaborate with.
But, the question still remains---how
can a user make sure that the nodes,
which have been granted the read permission,
will not stealthily leak the data, or
alter the application, or---even---how to ascertain
that they actually do anything
instead of returning an arbitrary result?

We believe such distrust is deep-seated in the nature of decentralization
and it might eventually become the major obstacle for the growth of the decentralized Internet.
In this paper,
we introduce a simple model to
systematically analyse different types of distrust.
And,
in order to tackle these problems,
we design \sys, a framework to mitigate the distrust in the decentralized Internet.

In our distrust model, multiple \userpods and one \textit{\dataconsumer}
together want to finish one \textit{task}.
An \textit{\executor} takes the data from \userpods
and the task from the \dataconsumer as inputs,
and produces the result.
All the participants mutually distrust each other.
There are concerns about the data counterfeit, data stealing, task stealing, execution corruption, and so on.

\sys is designed to mitigate the above distrust.
In its core are \textit{\tstep} and the \textit{\dexe}.
In \sys,
we split one task (e.g., a query, a machine learning training)
into steps, execute each step separately in a \tstep,
and chain the {\tstep}s chronologically as one \dexe.
Each \tstep consists of several trusted execution environments (TEE),
so that no data nor the execution leak outside the \dexe.

\sys's ethos is pragmatic.
It assumes TEE can provide both confidentiality and
integrity, which---in reality---is \textit{not} always true~\cite{xu2015controlled,l1tf}.
However, we argue that, in practice, the violations of such assumption
can be restricted by
(i) extra security techniques to solidify the TEE
and, more importantly,
(ii) thanks to the fine-granularity of each step,
the attacks would be economically insufficient.

\section{Distrust model}
\label{sec:model}

In this section, we are going to introduce our distrust model in the
decentralized Internet. We assume that the users' data have already been stored
in their \userpods. How to collect data into \userpods is out of the
topic of this paper.

\begin{figure}[h]
\centering
\includegraphics[width=0.47\textwidth]{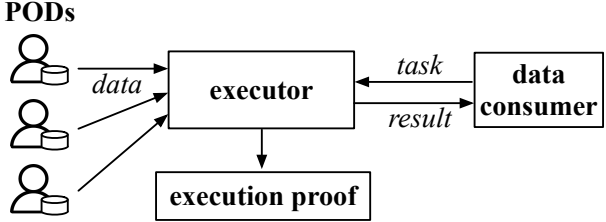}
\caption{The workflow of executing a task in the decentralized Internet.}
\label{fig:model}
\end{figure}

Figure~\ref{fig:model} depicts an abstract workflow of a task in the decentralized Internet.
A \textit{\dataconsumer} starts a \textit{task} which
includes a piece of code and a specification of what data should be its inputs.
Some \textit{\userpods} have the \textit{data} needed by this task
and (suppose) they allow such a task to access these data.

An \textit{\executor} takes the task from the \dataconsumer and
the corresponding \textit{data} from \textit{\userpods} as inputs,
executes the task's code,
and outputs the result to the \dataconsumer.
In addition, the \executor generates an \proof for
both sides to verify this procedure.
Note that the \executor is an abstract entity that might
locate in one or multiple nodes in the decentralized Internet.


In the above workflow, we have three participants with mutual distrust:
the \dataconsumer, \userpods, and the \executor.
\userpods are responsible for providing genuine data, but are fear of either \executor
or \dataconsumer stealing their data.
On the other hand, the \dataconsumer worries
that the \userpods may provide fake data,
and the task code are stolen or not faithfully executed by the \executor.
In a decentralized network, the participants
can be anyone in the world, hence they naturally distrust each other.

In the following,
we summarize the properties required
by the decentralized Internet
to mitigate the distrust.

\begin{itemize}

\item \textit{Data privacy}.
\userpods distrust the \dataconsumers that they may steal the data,
or the \executor might leak the data.
Data privacy requires that only a bounded amount of user data
can be revealed (i.e., differential privacy)
to the \dataconsumer and the \executor.

\item \textit{Data authenticity}.
\CF{\dataconsumers} worry that the \userpods may sabotage the task by sending
fake data as inputs.
Data authenticity requires that the input data
can be authenticated,
otherwise, should be explicitly tagged as alleged.

\item \textit{Task privacy}.
The task may contain some confidential logic of \dataconsumers'
(e.g., the architecture of a neural network, ranking algorithm)
that the \executor should not know.
Task privacy requires that the \executor learns nothing
about the task logic.

\item \textit{Execution integrity}.
\CF{\dataconsumers} concern that the result might be bogus,
as the task's code is not faithfully executed.
Execution integrity requires that the code runs as written
by the \executor.

\item \textit{Accountability}.
Inevitably, executing a task in a distributed setup will encounter
failures, errors, bugs, misconfigurations, and so on.
Accountability requires that, for one failed task, all the participants
are able to pinpoint and agree on where went wrong.

\item \textit{Traceability}.
When users need some processed data
(e.g., some statitics, a trained machine learning model),
they want to understand how the data were generated.
Traceability requires that one can find and verify the history of any piece of
data generated within the decentralized Internet.

\end{itemize}

\section{A draft solution: \sys}
\label{s:overview}

In order to fulfill the requires in previous section, we design \sys.
\sys is a framework built on top of a decentralized Internet,
where all the participating nodes are treated equally in a peer-to-peer manner.
Each node can play one or multiple roles---\userpods,
\dataconsumers, or the \executor---within each task.
And, \sys, as the framework, organizes and coordinates these roles
to finish the tasks.

In this section, we are going to introduce
the basic building block of \sys named \textit{\tstep} (\S\ref{subsec:t-step})
which provides execution integrity and task privacy;
then, we describe \textit{\dexe} (\S\ref{subsec:dexe}),
the key to
achieve the data privacy and data authenticity;
finally, we will justify our assumption in \S\ref{subsec:assumption}.

Due to the space limit, we have to skip other pieces of \sys
which are also indispensable.
In specific, we make the following assumptions, which should have been
achieved by other components of \sys.
\begin{myitemize}

\item We assume a fixed peer-to-peer network with all nodes equipped with TEE.

\item We assume all \userpods are willing to expose their data
(i.e., grant read permission)
under the guarantee of data privacy.

\item We assume that, for one task, both \userpods and the \dataconsumer
have agreed, beforehand, how to split this task into {\tstep}s.

\item We assume that \sys can always find the relevant data from \userpods
and qualified nodes as {\executor}s.

\end{myitemize}

\subsection{Properties of {\tstep}}
\label{subsec:t-step}

A {\tstep} is the basic execution unit in our system.
A task can be divided into multiple {\tstep}s.
Each {\tstep} has three properties:
first, it has only one entry point and one exit point for data;
second, the privacy of code and data within one {\tstep} is protected;
third, the execution of logic cannot be tampered with.

\begin{figure}[h]
\centering
\includegraphics[width=0.47\textwidth]{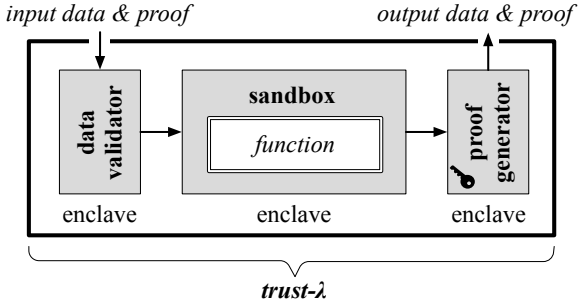}
\caption{The inner process of one {\tstep}. {\sys}-Box ensures only one entry for input and one exit for output.}
\label{fig:t-step}
\end{figure}

The properties of {\tstep} are enforced by a software and hardware co-design.
Figure~\ref{fig:t-step} shows the process of one \tstep.
Each \tstep is composed of three components: one data validator, one sandbox and one proof generator.
The data validator is the one component that accepts outside data.
It checks the input data using proof (described below), and then passes the data to the sandbox.
The sandbox ensures that the function running inside cannot send data out by disabling all I/O and restricting access to memory out of enclave.

The third component generates a proof of execution to show that the output data are indeed generated from the input data and algorithm.
A proof contains sufficient information of the data flow, but no sensitive data that may reveal user's personal information.
A typical proof only has hashes of the input data and the function.
The proof generator will sign the proof with its own private key.
The private key is initialized when the node joins the \sys network.

All of the three components are running within hardware enclaves, so that they can attest each other, offer attestation to outsiders and protect both integrity and privacy of the code and data.
They run on the same node so that their secure channels are based on shared memory buffers.


\subsection{\CF{\dexe}}
\label{subsec:dexe}

\begin{figure}[h]
\centering
\includegraphics[width=0.47\textwidth]{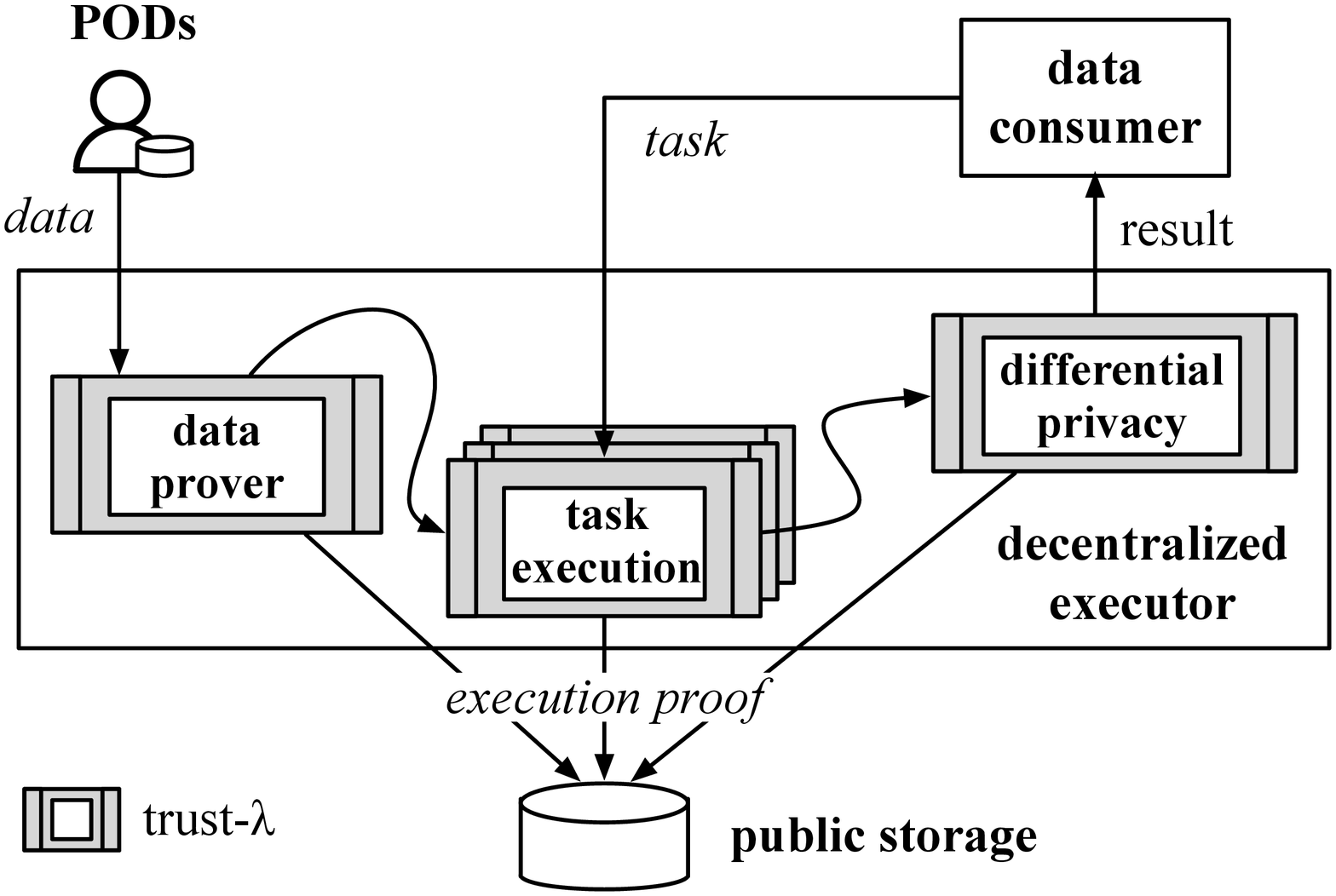}
  \caption{The architecture of \sys's \dexe.}
\label{fig:dexec}
\end{figure}

A \dexe is an instantiation of the \executor in a decentralized Internet,
which consists of a set of {\tstep}s.
Figure~\ref{fig:dexec} depicts the architecture of a \dexe.

In this example, the whole task has been split into
multiple {\tstep}s with different purposes:
\textit{data prover}, \textit{task execution}, and \textit{differential privacy}.
First, the input data flow from \userpods into the data prover,
in which the program authenticates the input data.
If valid, the data flows into the task execution to finish
the essential task the \dataconsumer provides.
The result, instead of directly returning to the \dataconsumer,
has to go through another differential privacy \tstep,
which guarantees that the sensitive personal data would not be leaked.
Finally, the result is delivered to the \dataconsumer.

All these procedures are executed within {\tstep}s,
which, because of \tstep's properties, are faithfully executed with confidentiality.
And, each of the \tstep commits an execution proof to a public storage,
which \userpods and the \dataconsumer can examine.

One \dexe usually has a chain of {\tstep}s.
%
The integrity of the \dexe is ensured by encryption and key management.
Specifically, the data consumer acts as a dispatcher.
It first recrutes several instances that runs the {\tstep} it needs, then sends the list to the data owner.
The data owner first attests each instance.
After that, it will generate keys and send them to each instance in a way that the output of one {\tstep} can only be decrypted by the next {\tstep}.



In the following, we're going to describe several common components
in a \dexe.

\heading{Data prover} is responsible for providing \textit{data authenticity} in the \dexe.
Depending on the type of data,
there are multiple ways to authenticate a piece of data.
If the data have been signed by the hardware (e.g., camera~\cite{friedman1993trustworthy}),
the validation can be as simple as verifying a signature.
Similar case applies to the data signed by their source organization
(e.g., health data~\cite{giakoumaki2006multiple}).

Another type of data, which have machine-checkable sources
(e.g., shopping record, search history),
can be authenticated by logging into the corresponding website
and verifying that the provided data entry does exist.
TEE technique~\cite{matetic2018delegatee} is able to guarantee
that user's login is secure,
as well as the validation is faithful.

Admittedly, there are cases that it is hard or even impossible
to authenticate the data. \sys requires that
these data should be tagged as ``alleged'' and the \dataconsumers
should take their own risks to trust these data.

\heading{Differential privacy} guarantees the \textit{data privacy} in \sys.
Differential privacy~\cite{dwork2014algorithmic}
is a statistical technique that aims to protect privacy of individual users
while still allowing statistical queries to these data.
Previous systems
demonstrate that differential privacy can be applied \textit{transparently}
to a lot of tasks
(e.g., machine learning~\cite{abadi2016deep},
statistic analysis~\cite{erlingsson2014rappor},
SQL query~\cite{johnson2018towards}).
Hence, \sys can leverage these systems to build our differential privacy \tstep.
Investigating what tasks can (or cannot) use differential privacy
and how to achieve them in practice is our future work.

\heading{Execution proofs} 
empower \sys with the \textit{accountability} and \textit{traceability}.
Any participant in one task is able to verify the whole chain of {\tstep}s
via examining (checking the signature and comparing the hashes)
the series of execution proofs on the public storage.
Since execution proofs are signed by {\tstep}s, they cannot be forged.

Specifically, \userpods can ensure that their data flow through
a differential privacy \tstep, so that data privacy is enforced.
And \dataconsumers can ascertain
that the input data have been authenticated by a data prover \tstep.
As for the data generated within \sys, anyone can track the history
of the processing and find the original data sources.

\subsection{Justify \sys's assumption}
\label{subsec:assumption}

\sys assumes that TEE can provide both execution integrity and confidentiality.
However, in reality, attacks, like side channel~\cite{xu2015controlled} and L1TF~\cite{l1tf},
may successfully steal data from enclaves.
In the following, we try to argue that
the violation of such assumption
can be restricted by the two schemes below.

One way to reduce the attacking surface,
besides applying Intel's latest patch,
is to restrict the software stack underneath enclaves with trusted hardware like TPM (Trusted Platform Module) and Intel TXT (Trusted eXecution Technology).
The execution nodes may have different security levels.
For examle, a node depending entirely on Intel SGX (Software Guard eXtension) is considered as mid-level security, which could be vulnerable to side channel attacks.
A more secure node can run a formally verified kernel and use SGX as well, and nothing else to minimize the attacking surface.

On the other hand, our methodology of dividing a task into multiple {\tstep} can
increase the cost of an attack,
as well as decrease its gain, making it economically insufficient.
Since {\tstep}s are isolated with each
other, the data leaked in a single {\tstep} can be very limited and an attacker
may need to compromise many nodes before achieving one sucessful attack.
Further, the functions that are considered as more important can be deployed to
the nodes with higher security level.


\section{Real-world use cases}
\label{s:scenario}

In this section, we are going to present several real-world use cases of \sys.

\subsection{Case 1: advertisements}

In this application scenario, advertisers are \dataconsumers.
They want to run their advertisement recommendation algorithms on users' \userpods
and deliver the advertisements about their products to the potential buyers.
Clearly, users do not want their data to be leaked to the advertisers,
and (perhaps) the advertisers also want their recommendation algorithms to be secret.

With \sys, an advertiser can send its advertisement to the target users
without learning who they are.
The advertiser can filter out
the non-target users by setting conditions in the task.
For example, as a video game seller, it can specify
that its advertisements will only send to those
who have a purchase record of a Nintendo Switch.
The advertiser can check the advertisement delivery by examining the execution proofs
committed by each \tstep.
\userpods can also make sure that their personal data (e.g., whether have a Nintendo Switch)
have not been leaked to the advertiser,
in the same way.

\subsection{Case 2: financial data query}

Some organizations (e.g., banks, financial companies)
and individuals have many financial records.
Other companies, researchers, or individuals may try to understand
some statistics about the overall financial status,
or the financial condition of a particular group of people.
However, nowadays, the data owners are unwilling to share their data,
and the queriers are afraid of that their queries will be leaked.

In this scenario, data providers need to ensure no data leakage,
and for the queriers, they need to keep their queries private.
With {\sys},
a PIR (Private Information Retrieve) \tstep can be inserted into the task
to protect the query processing from the data providers.
The query will be encrypted and executed by the \userpods of
data provider without revealing which data have been touched.
%
%
Since the data providers do not trust any query, they will
also leverage differential privacy to ensure no personal
information will be leaked through the result.


\subsection{Case 3: machine learning training} 

Machine learning brings convenience to people's lives: machine translation, autocorrect, auto captioning, etc. 
Most of these machine learning algorithms require training on substantial data. 
And some personalization functionalities require the peeking of user's own data. 
People have been forced to choose between convenience and privacy. 
\sys can make this dilemma obsolete. 

With \sys, people can opt in to 
volunteer their data for machine learning tasks, 
without the worry of losing privacy. 
These machine learning tasks can in turn be used for their own benefits, 
such as personal assistant, search personalization, etc.

Imagine a researcher wants to train a language model of modern day Americans, and 
she is interested in making use of the abundance of training data lying in people's\userpods.
In this case she would be a data consumer, who specifies the task and gets the word out. 
Some users that store their text messages in their \userpods may now sign up.
Data consumer starts the job in an enclave (consumer enclave), which updates the global model.
She then sends algorithm to user's \userpods which will run as \tstep.

The learning process involves several rounds of communcation between consumer enclave and user enclaves. 
In each round, the consumer enclave distributes the global model.
Each user enclave updates the global model with local data using stochastic gradient descent, encrypts parameters, and sends it back to the consumer enclave. 
The consumer enclave then collects the encrypted parameters, which it cannot decrypt. 
But it can aggregate the parameters and then decrypt the result. 
After a few rounds of updates, the global model stablizes, and training is done. 

\subsection{Case 4: secure polls and surveys}
Polls and surveys can be done on \sys.\
Survey conductors need integrity of the result,
and survey takers need discretion.

Say a company is trying to conduct a customer survey, to learn about how a product is received by the market.
In this case, it would be the data consumer.
Participating users can authenticate their legitimacy throught an identity verification protocal (e.g.,\ cryptographic proof of membership).
The demographic data can be filled out automatically using the data in \userpods.
Then the user can communicate their encrypted answers securely to the executor.
The executor then generates survey results, without revealing the raw data.
Eventually, a secure polling system may lead to a secure voting system.



\section{Related work}
\label{s:relwk}

\heading{Decentralized system.}
There are multiple decentralized systems~\cite{mansour2016demonstration,chajed2015amber,chajed2016oort}
designed to give back the control of data to users.
Solid~\cite{mansour2016demonstration}, a decentralized platform for social web applications,
defines a series of rich protocols and suggests the creation of a \textit{\userpod},
which is owned by individual users and stores their data.
Amber~\cite{chajed2015amber} and Oort~\cite{chajed2016oort}
provide global queries to efficiently collect and monitor relevant data created by all the users,
which enables cross-user data applications.
Instead of providing more functionalities in a decentralized setup,
\sys focuses on taming the distrust within these functionalities.

\heading{TEE.}
Many previous works leverage TEE to protect execution in untrusted environment~\cite{baumann2015shielding,arnautov2016scone,hunt2016ryoan,dinh2015m2r,schuster2015vc3}.
Ryoan~\cite{hunt2016ryoan} proposes a distributed sandbox mechanism that enables a user to process her private data (e.g., gene) by multiple untrusted algorithms on different platforms.
Our design is inspired by Ryoan but targets a different problem: the data owner is not the consumer, which leads to a totally different threat model.
For example, Ryoan does not need to consider the fake data problem or privacy leakage between data owner and consumer, while \sys considers both.

\heading{Cryptographic tools.}
There is another thread of systems~\cite{gupta2017pretzel,gilad2016cryptonets,setty2018proving,angel2016unobservable}
that leverage model cryptographic tools,
such as FHE (Fully-Homomorphic Encryption),
MPC (secure Multi-Party Computation),
PIR (Private Information Retrieval),
to \textit{completely} get rid of the trust among participants.
\sys is complementary to them.
Meanwhile, these methods 
can be implemented within \tstep as a replacement of TEE.
However, as far as we know, adopting these techniques
usually imposes significant overheads, hence we choose to use TEE as our current option.

\heading{Privacy-preserving machine learning.}
Federated learning \cite{mcmahan2016communication} is a learning scheme that allows a model to be trained on remote and decentralized data. 
This protects privacy to some extend because data stay where they are. 
Cryptographic tools have been used to prevent adversarial inference of individual data \cite{bonawitz2017practical}.
It is worth mentioning that multiple attacks on federated learning have been proposed \cite{hitaj2017deep}.

Another line of efforts has been on acheving differential privacy with random noise mechanisms \cite{dwork2009differential, abadi2016deep}. 
They are pertinent on particular machine learning algorithms, and proved differential privacy parameters. 
\sys can seamlessly accomodate these designs.

\section{Conclusion}
\label{s:conclusion}

Decentralization is not just about data (e.g., PODs) and execution (e.g., distributed computing), it is also about \textbf{trust}.
The decoupling of three roles: data owner, data consumer and executor, brings new challenges to the trust model as well as the system design.
\sys proposes a decentralized execution framework, where each step of execution goes with proofs of data and code.
Multiple steps can be connected to run various tasks with code and data from different roles.
Currently, \sys leverages hardware TEE to achieve better practicability.
It can also use cryptographic tools including MPC and FHE according to different scenarios.

\balance
\bibliographystyle{abbrvnat}
\footnotesize
\setlength{\bibsep}{3pt}
\bibliography{p,conf}
\end{document}